\documentclass[letterpaper, 10 pt, conference]{ieeeconf}  % Comment this line out
\IEEEoverridecommandlockouts                              % This command is only
\usepackage{graphicx} % for pdf, bitmapped graphics files
\usepackage{booktabs} % For formal tables
\usepackage{todonotes}
\usepackage{cite}
\usepackage{hyperref}
\hypersetup{
    colorlinks=true,
    linkcolor=black,
    filecolor=black,      
    urlcolor=blue,
    }
    
\usepackage{multicol}
\usepackage{amsmath}
\usepackage{amsfonts}
\usepackage{xcolor}
\newcommand{\eat}[1]{}

\usepackage{subfig}
\usepackage{graphicx}

\title{\LARGE \bf
Zero-Shot Autonomous Vehicle Policy Transfer: From Simulation to Real-World via Adversarial Learning
}

\author{Behdad Chalaki, Logan E. Beaver, Ben Remer, Kathy Jang, Eugene Vinitsky,\\ Alexandre M. Bayen, Andreas A. Malikopoulos 
\thanks{This research was funded in part by the Delaware Energy Institute (DEI) and in part by an NSF Graduate Research Fellowship. AWS credits and funding were provided by an Amazon Machine Learning Research award.}
\thanks{Behdad Chalaki, Logan E. Beaver, Ben Remer, and Andreas A. Malikopoulos are with the Department of Mechanical Engineering, University of Delaware, Newark, DE, USA (email: bchalaki@udel.edu, lebeaver@udel.edu, bremer@udel.edu, andreas@udel.edu).}
\thanks{Kathy Jang and Alexandre Bayen are with the Department of Electrical Engineering and Computer Sciences, and Eugene Vinitsky is with the Department of Mechanical Engineering University of California, Berkeley, CA, USA (email: kathyjang@berkeley.edu, bayen@berkeley.edu, evinitsky@berkeley.edu).}
}

\begin{document}

\maketitle %\todo{include IEEE membership status on author list?}
\thispagestyle{empty}
\pagestyle{empty}

%%%%%%%%%%%%%%%%%%%%%%%%%%%%%%%%%%%%%%%%%%%%%%%%%%%%%%%%%%%%%%%%%%%%%%%%%%%%%%%%
\begin{abstract}

In this article, we demonstrate a zero-shot transfer of an autonomous driving policy from simulation to University of Delaware's scaled smart city with adversarial multi-agent reinforcement learning, in which an adversary attempts to decrease the net reward by perturbing both the inputs and outputs of the autonomous vehicles during training. We train the autonomous vehicles to coordinate with each other while crossing a roundabout in the presence of an adversary in simulation. The adversarial policy successfully reproduces the simulated behavior and incidentally outperforms, in terms of travel time, both a human-driving baseline and adversary-free trained policies. Finally, we demonstrate that the addition of adversarial training considerably improves the performance \eat{stability and robustness} of the policies after transfer to the real world compared to Gaussian noise injection.% \\Supplementary information and videos can be found at: ~~~~ \url{https://sites.google.com/view/ud-ids-lab/arlv}
\end{abstract}

%%%%%%%%%%%%%%%%%%%%%%%%%%%%%%%%%%%%%%%%%%%%%%%%%%%%%%%%%%%%%%%%%%%%%%%%%%%%%%%%

\section{Introduction}

%\todoBehdad{\textcolor{blue}{ Blue for new changes}, \textcolor{gray}{gray for parts to be removed}}

\eat{
\textcolor{gray}{
We are witnessing an increasing integration of our energy, transportation, and cyber network systems, which coupled with human interactions, is giving rise to a new level of complexity in mobility systems. As these increasingly complex mobility system emerge, new approaches are needed to optimize the efficiency of the transportation system by considering the 
interactions between vehicles at different transportation scenarios, e.g., intersections, merging roadways, roundabouts, and speed reduction zones. These scenarios, along with the drivers' responses to various disturbances, are the primary sources of bottlenecks that contribute to traffic congestion \cite{Margiotta2011}.
}}
In $2015$, commuters in the US spent an estimated $6.9$ billion additional hours waiting in congestion, resulting in an extra $3.1$ billion gallons of fuel, costing an estimated \$$160$ billion \cite{Schrank2015}. 
An automated transportation system \cite{Zhao2019} can alleviate congestion, reduce energy use and emissions, and improve safety by increasing traffic flow.
The use of connected and automated vehicles (CAVs) can transition our current transportation networks into energy-efficient mobility systems. Introducing CAVs into the transportation system allows vehicles to make
better operational decisions, leading to significant reductions of energy consumption, greenhouse gas emissions, and travel delays along with improvements to passenger safety \cite{Wadud2016}.
\eat{\textcolor{gray}{Additionally, the generation of massive amounts of vehicle data creates significant opportunities to develop optimization algorithms for controlling large-scale behaviors for entire urban systems.
}}

Several efforts have been reported in the literature towards coordinating CAVs to reduce spatial and temporal speed variations of individual vehicles. % throughout the transportation network. 
These variations can be introduced by breaking events, or due to the structure of the road network, e.g., intersections \cite{Lee2012, rakha2011eco, Malikopoulos2017}, cooperative merging, and speed reduction zones \cite{malikopoulos2018optimal}. One of the earliest efforts in this direction was proposed by Athans \cite{Athans1969} to efficiently and safely coordinate merging behaviors as a step to avoid congestion. With an eye toward near-future CAV deployment, several recent studies have explored the traffic and energy implications of partial penetration of CAVs under different transportation scenarios, e.g., \cite{Rios2018, Malikopoulos2016a, guanetti2018, wang2018r,guanetti2018control}.

While classical control is an effective method for some traffic control tasks, the complexity and sheer problem size of autonomous driving in mixed-traffic scenarios makes it a notoriously difficult problem to address. 
In recent years, deep reinforcement learning (RL) has emerged as an alternative method for traffic control. RL is recognized for its ability to solve data-rich, complex problems such as robotic skills learning \cite{gu2017deep}, to larger and more complicated problems such as learning winning strategies for Go \cite{silver2017mastering} or StarCraft II~\cite{deepmind_2019}. Deep RL is capable of handling highly complex behavior-based problems, and thus naturally extends to traffic control.
The results from the ring road experiments that Stern and Sugiyama \cite{stern2017dissipation, sugiyama2008traffic} demonstrated their policies have also been achieved via RL methods \cite{wu2017emergent}. AVs controlled with RL-trained policies have been further used to demonstrate their traffic-smoothing capabilities in simple traffic scenarios such as figure eight road networks \cite{wu2017emergent}, intersections \cite{vinitsky2018benchmarks} and roundabouts \cite{Jang2019SimulationVehicles}, and can also replicate traffic light ramp metering behavior \cite{belletti2017expert}. \eat{\textcolor{gray}{Eco-driving in multiple signalized intersections have enabled CAVs to reduce fuel consumption by $50-57\%$ in compare with the modified intelligent driver model while maintaining arrival time \cite{sun2018robust}.}} 
Indeed, real-world evaluation and validation of control techniques under a variety of traffic scenarios
is a necessary task.
\eat{\textcolor{gray}{
to ensure the theoretical guarantees of CAV control policies.}}

The contributions of this article are: (1) the introduction of Gaussian single-agent noise and adversarial multi-agent noise to learn traffic control behavior for an automated vehicle;
    (2) a comparison performance with noise injected into the action space, state space, and both;
    (3) the demonstration of real-world disturbances leading to poor performance and crashes for some training methods, and
    (4) experimental demonstration of how autonomous vehicles can improve performance in a mixed-traffic system.

The remainder of this article is organized as follows. In Section \ref{sec:background}, we provide background information on reinforcement learning, car-following models, the \textit{Flow} framework, and the experimental testbed. In Section \ref{sec:problemFormulation}, we introduce the mixed-traffic roundabout problem and the implementation of the RL framework. In Section \ref{sec:sim}, we present the simulation results. In Section \ref{sec:experiment}, we discuss the policy transfer process along with the experimental results. Finally, we draw concluding remarks in Section \ref{sec:conclusions}.

\section{Background} \label{sec:background}
\subsection{Deep Reinforcement Learning}
RL is a subset of machine learning which studies how an \textit{agent} can take \textit{actions} in an \textit{environment} to maximize its expected cumulative reward. The environment in which RL trains its agent is modeled as a Markov decision processes~\cite{bellman1957markovian}, which is the model that we used for all experiments in this article.
A finite-horizon, discounted Markov decision process is defined by the tuple $(\mathcal{S}, \mathcal{A}, P, r, \rho_0, \gamma, T)$, where
$\mathcal{S} \subseteq \mathbb{R}^n$ is an $n$-dimensional set of states; $\mathcal{A} \subseteq \mathbb{R}^m$ is an $m$-dimensional set of actions, $P: \mathcal{S} \times \mathcal{A} \times \mathcal{S} \to \mathbb{R}_{\geq 0}$ describes the transitional probability of moving from one state $s$ to another state $s'$ given an action $a$; $r : \mathcal{S} \times \mathcal{A} \to \mathbb{R}$ is the reward function; $\rho_0: \mathcal{S} \to \mathbb{R}_{\geq 0}$ is the probability distribution over start states; $\gamma \in (0, 1]$ is the discount factor; and $T$ is the horizon.

\eat{\textcolor{gray}{
In RL, an agent iteratively receives sensory information describing its environment in the form of $\mathcal{S}$. Based on $\mathcal{S}$, the agent decides on what actions $\mathcal{A}$ to take to advance to the following state $s'$. These actions are chosen from a stochastic \emph{policy} $\pi: \mathcal{S} \to \mathcal{A}$. The goal of RL is to learn an optimal policy $\pi^*: \mathcal{S} \to \mathcal{A}$ by maximizing $R = \mathbb{E} \left[\sum_{t=0}^T \gamma^t r_t \right]$, where $r_t$ is the reward received at time $t$. This goal learns the best actions to take from any given state to maximize the expected cumulative reward.
}}

Deep RL is a form of RL which parameterizes the policy $\pi: \mathcal{S}\to\mathcal{A}$ with the weights of a neural net. The neural net consists of an input layer, which accepts state inputs $s \in \mathcal{S}$; an output layer, which returns actions $a \in \mathcal{A}$; and hidden layers, consisting of affine transformations and non-linear activation functions. The flexibility that hidden layers provide neural nets with the possibility of being \textit{universal function approximators}, and enables RL policies to express complex functions.

\subsection{Policy Gradient Algorithms}
There are a number of algorithms that exist for deriving an optimal RL policy $\pi^*$. For the experiments in this article, $\pi^*$ is learned via proximal policy optimization (PPO)~\cite{schulman2017proximal}, a widely-used policy gradient algorithm. Policy gradient algorithms operate in the policy space by computing an estimate of the gradient of the expected reward $\nabla_{\theta} R = \nabla_{\theta} \mathbb{E}\left[\sum_{t=0}^T \gamma^t r_t \right]$, where $\theta$ is the parameters of the policy. The policy is then updated by performing gradient ascent methods to update $\theta$.

In this article, we use PPO as the algorithm for the two types of experiments described in Sec. \ref{sec:problemFormulation}, Gaussian single-agent and adversarial multi-agent. PPO uses a clipped surrogate objective to perform each policy update, giving it stability and reliability similar to trust-region methods such as TRPO \cite{schulman2015trust}.

\subsection{Car Following Models} \label{sec:carFollowing}
We use the intelligent driver model (IDM)~\cite{Treiber2000} to model  human driving dynamics. IDM is a time-continuous microscopic car-following model which is widely used in vehicle motion modeling. Using the IDM, the acceleration for vehicle $\alpha$ is a function of its distance to the preceding vehicle, or the headway $s_\alpha$, the vehicle' own velocity $v_\alpha$, and relative velocity, $\Delta v_\alpha$, namely,

\begin{equation} \label{eq:idm}
a_{\text{IDM}} = \frac{dv_\alpha}{dt} = a \bigg[ 1 - \bigg( \frac{v_\alpha}{v_0} \bigg)^\delta - \bigg( \frac{s^*(v_\alpha,\Delta v_\alpha)}{s_\alpha} \bigg)^2 \bigg],
\end{equation}
where $s^*$ is the desired headway of the vehicle, % and is given
\begin{equation} \label{eq:s_star}
s^*(v_\alpha,\Delta v_\alpha) = s_0 + \max \bigg( 0, v_\alpha T + \frac{v_\alpha \Delta v_\alpha}{2 \sqrt{ab}} \bigg),
\end{equation}
where $s_0, v_0, T, \delta, a, b$ are known parameters. We describe these parameters and the values used in our simulation in Section ~\ref{sec:sim}.

\subsection{Flow}
For training the RL policies in this article, we use \textit{Flow} \cite{wu2017flow}, an open-source framework for interfacing RL libraries such as RLlib \cite{liang2017ray}, Stable Baselines, and rllab \cite{duan2016benchmarking} with traffic simulators such as SUMO \cite{SUMO2012} or Aimsun. \textit{Flow} enables the ability to design and implement RL solutions for a flexible, wide variety of traffic-oriented scenarios. RL environments built using \textit{Flow} are compatible with OpenAI Gym \cite{Brockman2016} and as such, support training with most RL algorithms. \textit{Flow} also supports large-scale, distributed computing solutions via AWS EC2 \footnote{For further information on Flow, we refer readers to view the \textit{Flow} Github page, website, or article, respectively listed here. Github: \textcolor{blue}{\url{https://github.com/flow-project/flow}}, website:  \textcolor{blue}{\url{https://flow-project.github.io/}}, paper: \cite{wu2017flow}.}.

\section{Problem Formulation} \label{sec:problemFormulation}

\begin{figure}[ht]
    \centering
    \includegraphics[width=0.6\linewidth]{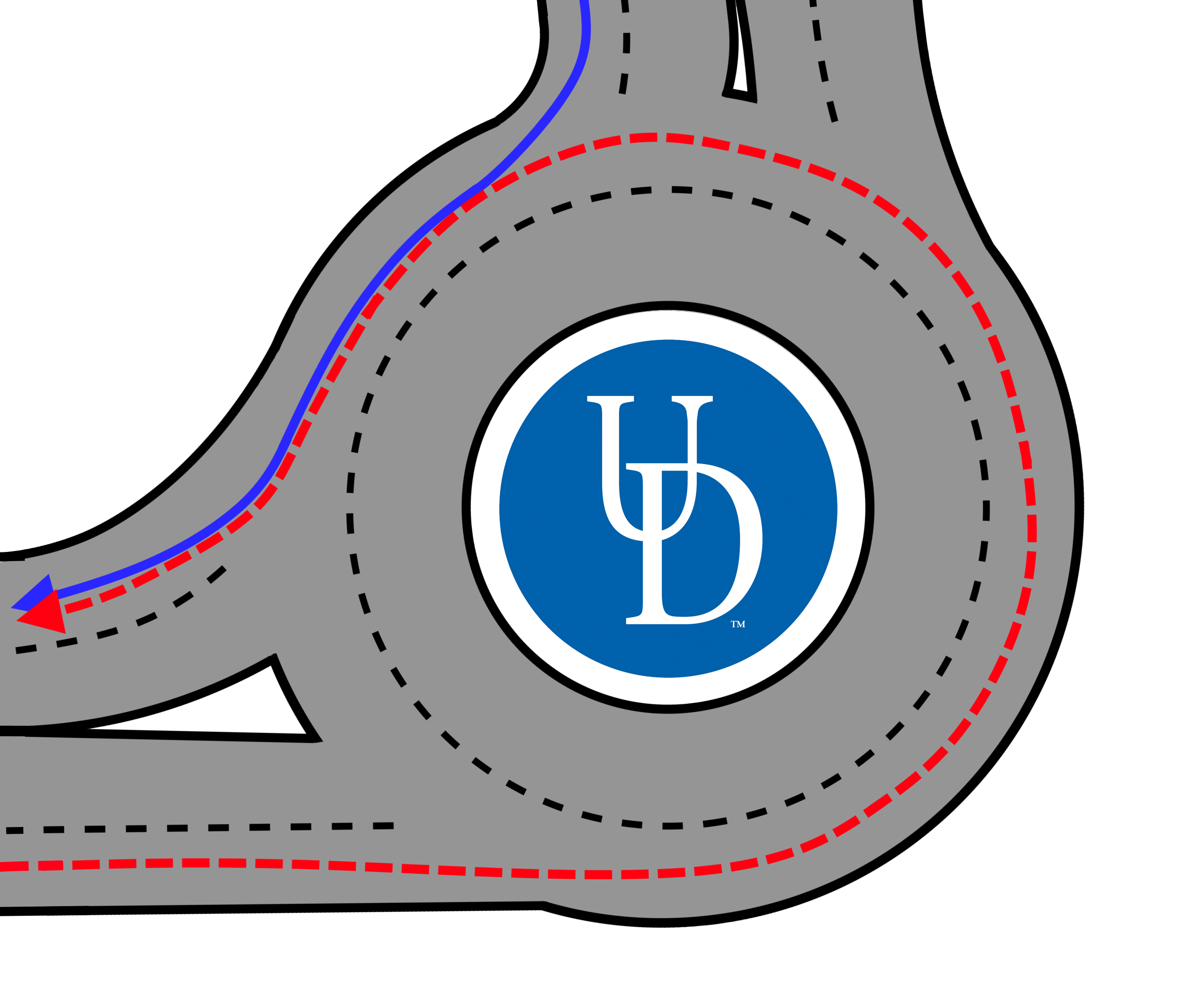}
    \caption{The routes taken by the northern (solid blue) and western (dashed red) groups through the roundabout.}
    \label{fig:udsscRoutes}
\end{figure}

To demonstrate the viability of autonomous RL vehicles in reducing congestion in mixed traffic, we implemented the scenario shown in Fig. \ref{fig:udsscRoutes}. In this scenario, two groups of vehicles enter the roundabout stochastically, one at the northern end and one at the western end. In what follows, we refer to the vehicles entering from the north entry as the \textit{northern group}, and to the vehicles entering from the west entry as the \textit{western group}.

The baseline scenario consists of homogeneous human-driven vehicles using the IDM controller \eqref{eq:idm}. The baseline is designed such that vehicles approaching the roundabout from either direction will clash at the roundabout. This results in vehicles at the northern entrance yielding to roundabout traffic, resulting in significant travel delays. The RL scenario puts an autonomous vehicle at the head of each group, which can be used to control and smooth the interaction between vehicles; these mixed experiments correspond to a $15\%-50\%$ mixture of autonomous and human-driven vehicles.

\subsection{Reinforcement Learning Structure}
We categorize two sets of RL experiments that are used and compared in this article. We will refer to them as:
    \emph{Gaussian single-agent}: A single-agent policy trained with Gaussian noise injected into the state and action space.
    \emph{Adversarial multi-agent}: A multi-agent policy trained wherein a second agent provides selective adversarial noise to the learning agent.
%\end{enumerate}
We discuss the particulars of these two methods in Sections  \ref{sec:gaussian-noise} and \ref{sec:adversarial-noise}. In this article, we deploy seven RL-trained policies, one of which is single-agent with no noise, three of which are Gaussian single-agent and the other three of which are adversarial multi-agent. All experiments follow the same setup. Inflows of stochastic length emerge at the northern and western ends of the roundabout. The size of the northern group will range from $2$ to $5$ cars, while the size of the western group ranges from $2$ to $8$. The length of these inflows will remain static across each rollout, and are randomly selected from a uniform distribution at the beginning of each new rollout.

\subsubsection{Action Space}

The actions are applied from a $2$-dimensional acceleration vector, in which the first element is used to control the AV leading the northern group, and the second is used to control the AV leading the western group. If the AV has left the experiment, that element of the action vector is discarded.

\subsubsection{State Space}
\label{sec:state-space}

The state space conveys the following information about the environment to the agent:
    the position, velocity, tailway, and headway of each AV and each vehicle in the roundabout, the distance from the roundabout entrances to the $6$ closest vehicles, the velocities of the $6$ closest vehicles to each roundabout entrance, the number of vehicles queued at each entrance, and the lengths of each inflow.
\eat{\textcolor{gray}{
\begin{itemize}
    \item The positions of the AVs.
    \item The velocities of the AVs.
    \item The distances from the roundabout of the $6$ closest vehicles to the roundabout for both roundabout entryways.
    \item The velocities of the $6$ closest vehicles to the roundabout for both roundabout entryways.
    \item Tailway and headway (i.e., distances to the leading and following vehicles) of vehicles from both AVs.
    \item Length of the number of vehicles waiting to enter the roundabout for both roundabout entryways.
    \item The positions and velocities of all vehicles in the roundabout.
    \item Lengths of the incoming inflows. 
\end{itemize}
}}All elements of the state space are normalized. The state space was designed with real-world implementation in mind and could contain any environmental factors that the simulation supports. As such, it is partially-observable to support modern sensing capabilities. All of these observations are reasonably selected and could be emulated in the physical world using sensing tools such as induction loops, camera sensing systems, and speedometers. 
\subsubsection{Reward Function}
The reward function used for all experiments minimizes delay and applies penalties for standstill velocities, near-standstill velocities, jerky driving, and speeding, i.e.,

\begin{equation}
r_t = 2 \cdot \frac{\max\left({v_\text{max}\sqrt{n} - \sqrt{\sum_{i=1}^{n}(v_{i, t}-v_\text{max})^2}}, 0\right)}{v_\text{max}\sqrt{n}} - p,
\end{equation}
\begin{equation}\label{eq:p}
    p = p_s + p_p + p_j + p_v.
\end{equation}
where $n$ is the total number of vehicles, $p$ is the sum of four different penalty functions, $p_s$ is a penalty for vehicles traveling at zero velocity, designed to discourage standstill; $p_p$ penalizes vehicles traveling below $0.2 m/s$, which discourages the algorithm from learning an RL policy which substitutes extremely low velocities to circumvent the zero-velocity penalty; $p_j$ discourages jerky driving by maintaining a dynamic queue containing the last $10$ actions and penalizing the variance of these actions; and $p_v$ penalizes speeding.

\subsubsection{Gaussian single-agent noise}
\label{sec:gaussian-noise}
Injecting noise directly to the state and action space has been shown to aid with transfer from simulation to real-world testbeds \cite{Jang2019SimulationVehicles,tobin2017domain}. In this method, which applies to three of the policies we deployed, each element of the state space was perturbed by a random number selected from a Gaussian distribution. Only two elements describing the length of the inflows approaching the merge were left unperturbed. Elements of the state space corresponding to positioning on the merge edge were perturbed from a Gaussian distribution with a standard deviation of $0.05$. 
For elements corresponding to absolute positioning, the standard deviation was 0.02. 
All other elements used a standard deviation of $0.1$. 
These values were selected to set reasonable bounds for the degree of perturbation in the real world. Each element of the action space was perturbed by a random number selected from a zero mean Gaussian distribution with $0.5$ standard deviation. 

\subsubsection{Adversarial multi-agent noise}
\label{sec:adversarial-noise}
For the other $3$ policies, we use a form of adversarial training to yield a policy resistant to noise \cite{pinto2017robust}. This is a form of multi-agent RL, in which two policies are learned. Adversarial training pits two agents against each other in a zero-sum game. The first is structurally the same as the agent which is trained in the previous $4$ policies. The second, \textit{adversarial} agent has a reward function that is the negative of the first agent's reward; in other words, it is incentivized by the first agent's failure. The adversarial agent can attempt to lower the agent reward by perturbing elements of the action and state space of the first agent. 

The adversarial agent's action space is a 1-dimensional vector of length $22$, composed of perturbation values bound by $[-1, 1]$. The first two elements of the adversarial action space are used to perturb the action space of the original agent's action space. Adversarial action perturbations are scaled by $0.1$. Combining adversarial training with selective randomization, the adversarial agent has access to perturb a subset of the original agent's state space. The remaining $20$ elements of the adversarial agent's action space are used to perturb $20$ selective elements of the original agent's state space. Both the adversarial action and state perturbations are scaled down by $0.1$. The selected elements that the adversary can perturb are the observed positions and velocities of both controlled AVs in the system and the observed distances of vehicles from the merge points.

\section{Simulation Framework} \label{sec:sim}

\subsection{Car Following Parameters}
As introduced in Section ~\ref{sec:carFollowing}, the human-driven vehicles in these simulations are controlled via IDM. Accelerations are provided to the vehicles via 
\eqref{eq:idm} and \eqref{eq:s_star}. Within these equations, $s_0$ is the minimum spacing or minimum desired net distance from the vehicle in front of it, $v_0$ is the desired velocity, $T$ is the desired time headway, $\delta$ is an acceleration exponent, $a$ is the maximum vehicle acceleration, and $b$ is the comfortable braking deceleration. 

Human-driven vehicles in the system operate using SUMO's built-in IDM controllers, which allows customization to the parameters described above. Standard values for these parameters as well as a detailed discussion on the experiments producing these values can be found in \cite{Treiber2000}. In these experiments, the parameters of the IDM controllers are defined to be $T=1~\text{s}$ , $a=1~\text{m}/\text{s}^2$ , $b=1.5~\text{m}/\text{s}^2$, $\delta=4$, $s_0=2~\text{m}$, $v_0=30~\text{m}/\text{s}$. A noise parameter $0.1$ was used to perturb the acceleration of the IDM vehicles.

\eat{\textcolor{blue}{
To model the natural variance in human driving behavior, we supply stochasticity to SUMO's IDM controller by setting noise to $0.1$. In doing this, for every human-vehicle acceleration, SUMO samples from a zero-mean Gaussian distribution with a variance indicated by the value of noise, and perturbs the acceleration by this amount.
}}

Environment parameters in the simulation were set to match the physical constraints of the experimental testbed. These include: a maximum acceleration of $1~ \text{m}/\text{s}^2$, a maximum deceleration of $-3~ \text{m}/\text{s}^2$, and a maximum velocity of $8~  \text{m}/\text{s}$. The timestep of the system is set to $1~\text{s}$. 

\subsection{Algorithm/Simulation Details}
We ran experiments with a discount factor of $0.999$, a trust-region size of $0.01$, a batch size of $20000$, a horizon of $500$ seconds, and trained over $100$ iterations. The controller is a neural network, a \emph{Gaussian multi-layer perceptron} with a tanh non-linearity, and hidden sizes of $(100, 50, 25)$. The choice of neural network non-linearities, size, and type were picked based on traffic controllers developed in \cite{vinitsky2018benchmarks}. The states are normalized so that they are between $0$ and $1$ by dividing each state by its maximum possible value. The agent actions are clipped to be between $-3$ and $1$. Both normalization and clipping occur after the noise is added to the system so that the bounds are properly respected. 
%\subsection{Code Reproducibility}
The following codebases are needed to reproduce the results of our work. \emph{Flow}\footnote{\textcolor{blue}{\url{https://github.com/flow-project/flow}}.}, \emph{SUMO}\footnote{\textcolor{blue}{\url{https://github.com/eclipse/sumo}} at commit number \textbf{1d4338ab80}.} and the version of \emph{RLlib}\footnote{\textcolor{blue}{\url{https://github.com/flow-project/ray/tree/ray_master}} at commit number \textbf{ce606a9}.} used for the RL algorithms is available on GitHub.

\subsection{Simulation Results} 
\begin{figure}[ht]
\centering
\includegraphics[width=0.45\textwidth]{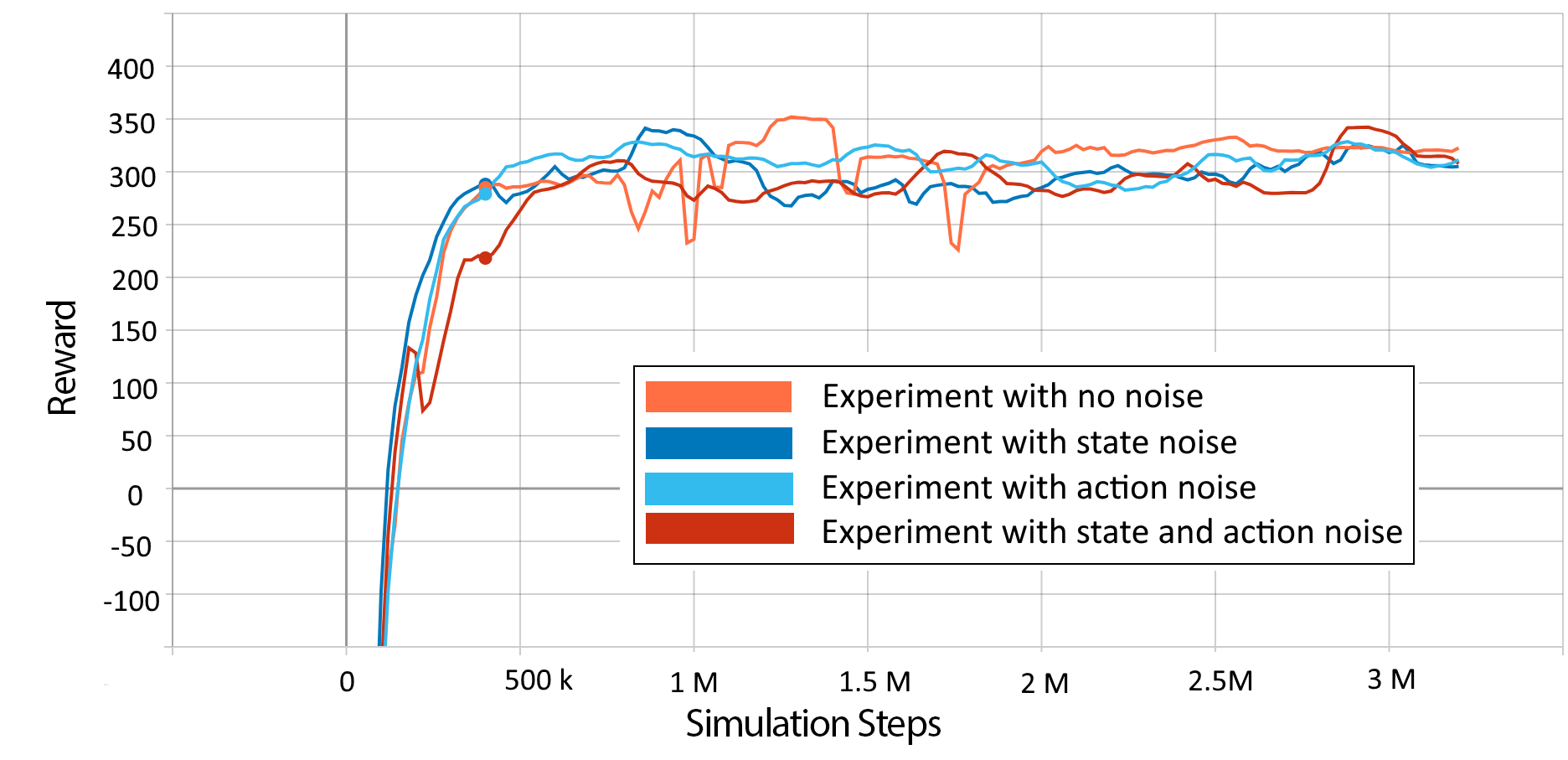}
\caption{Convergence of the RL reward curves of the 3 Gaussian %single-agent e
experiments and the noiseless policy.} % single-agent policy trained sans noise. The curves are superimposed. }
\label{fig:rewards}
\end{figure}

\begin{figure*}[ht]
\begin{multicols}{3}
  \includegraphics[width=.7\linewidth]{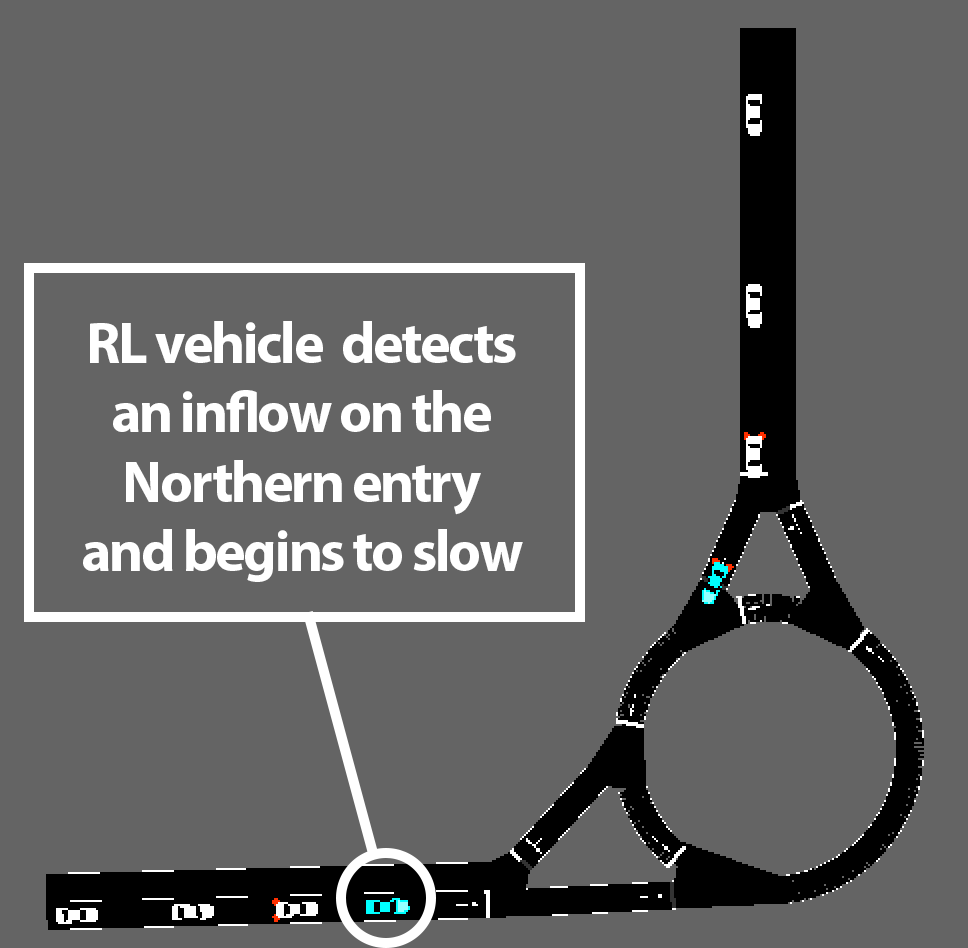} \par 
  \includegraphics[width=.7\linewidth]{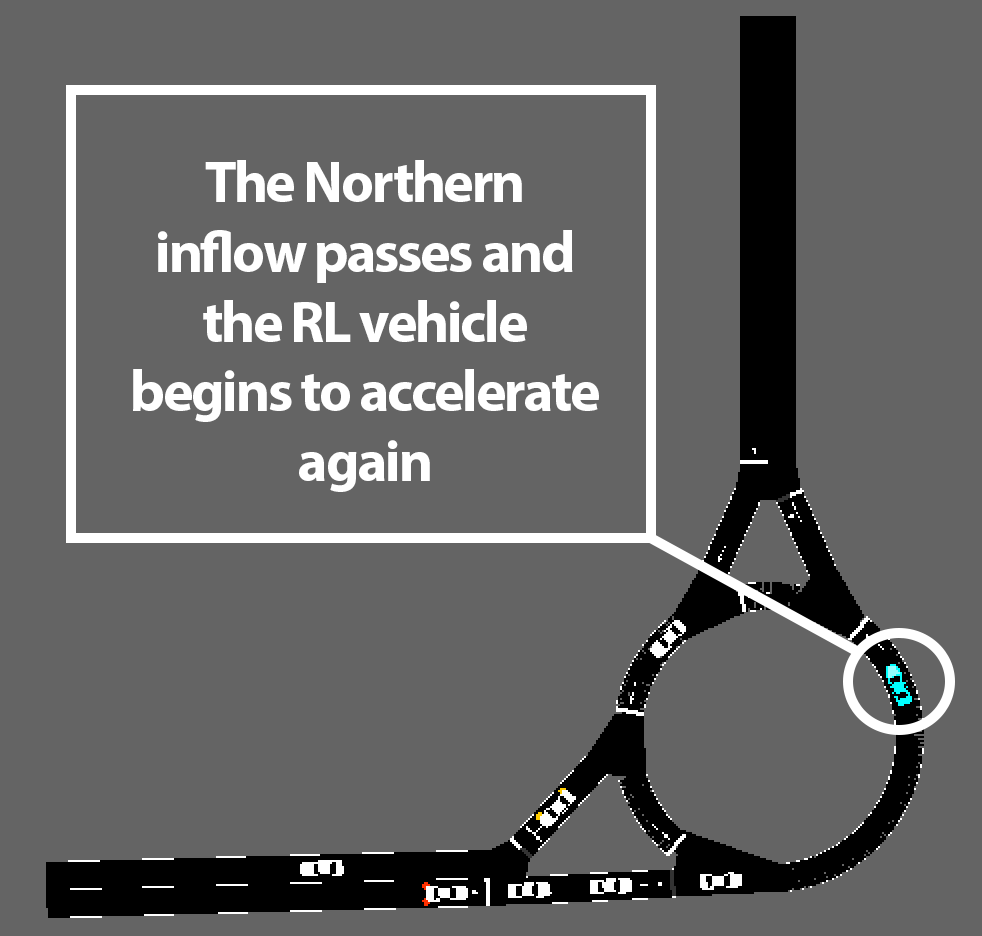}\par 
  \includegraphics[width=.7\linewidth]{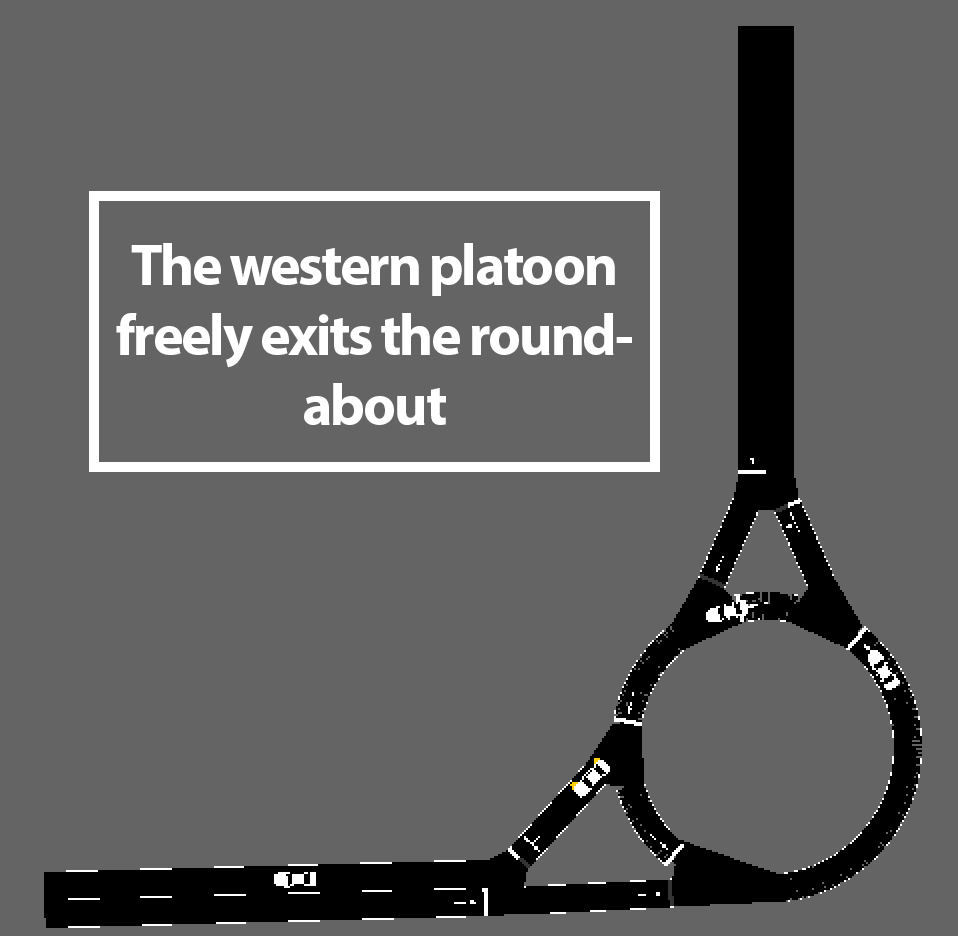} \par
\end{multicols}
\caption{Two RL-controlled AVs trained with adversarial multi-agent noise demonstrate emergent ramp metering behavior.}% in this series of images.}%, followed by an image of the baseline where vehicles clash. \textbf{First}: RL vehicle slows down in anticipation of a sufficiently small inflow from the north. \textbf{Second}: The northern inflow passes through the roundabout at high velocity. \textbf{Third}: The western group exits the roundabout at high velocity. \textbf{Fourth}: The baseline in which all vehicles are human-driven results in queues and clashing groups.}
\label{fig:roundabout-progression}
\end{figure*}
The reward curves of the Gaussian single-agent experiments are displayed in Fig. \ref{fig:rewards}. These include the curves of the $3$ experiments, which are trained with Gaussian noise injection, as well as one trained without any noise. 
In both the Gaussian single-agent and adversarial multi-agent experiments, the policy learns a classic form of traffic control: ramp-metering, in which one group of vehicles slows down to allow for another group of vehicles to pass. Despite the varying length of inflows from the two entries, policies consistently converge to demonstrate ramp-metering.

\section{Experimental Deployment} \label{sec:experiment}

\subsection{The University of Delaware's Scaled Smart City}
University of Delaware's Scaled Smart City (UDSSC) % (Fig. \ref{fig:udssc}) 
is a $1$:$25$ scale testbed designed to replicate real-world traffic scenarios and implement cutting-edge control technologies in a safe and scaled environment. UDSSC is a fully integrated smart city, which can be used to validate the efficiency of control and learning algorithms, including their performance on physical hardware. UDSSC utilizes high-end computers and a VICON motion capture system to simulate a variety of control strategies with as many as $35$ scaled CAVs. For further information on the capabilities and features of the UDSSC, see \cite{Beaver2020DemonstrationCity}.
\eat{
\begin{figure}
    \centering
    \includegraphics[width=0.45\textwidth]{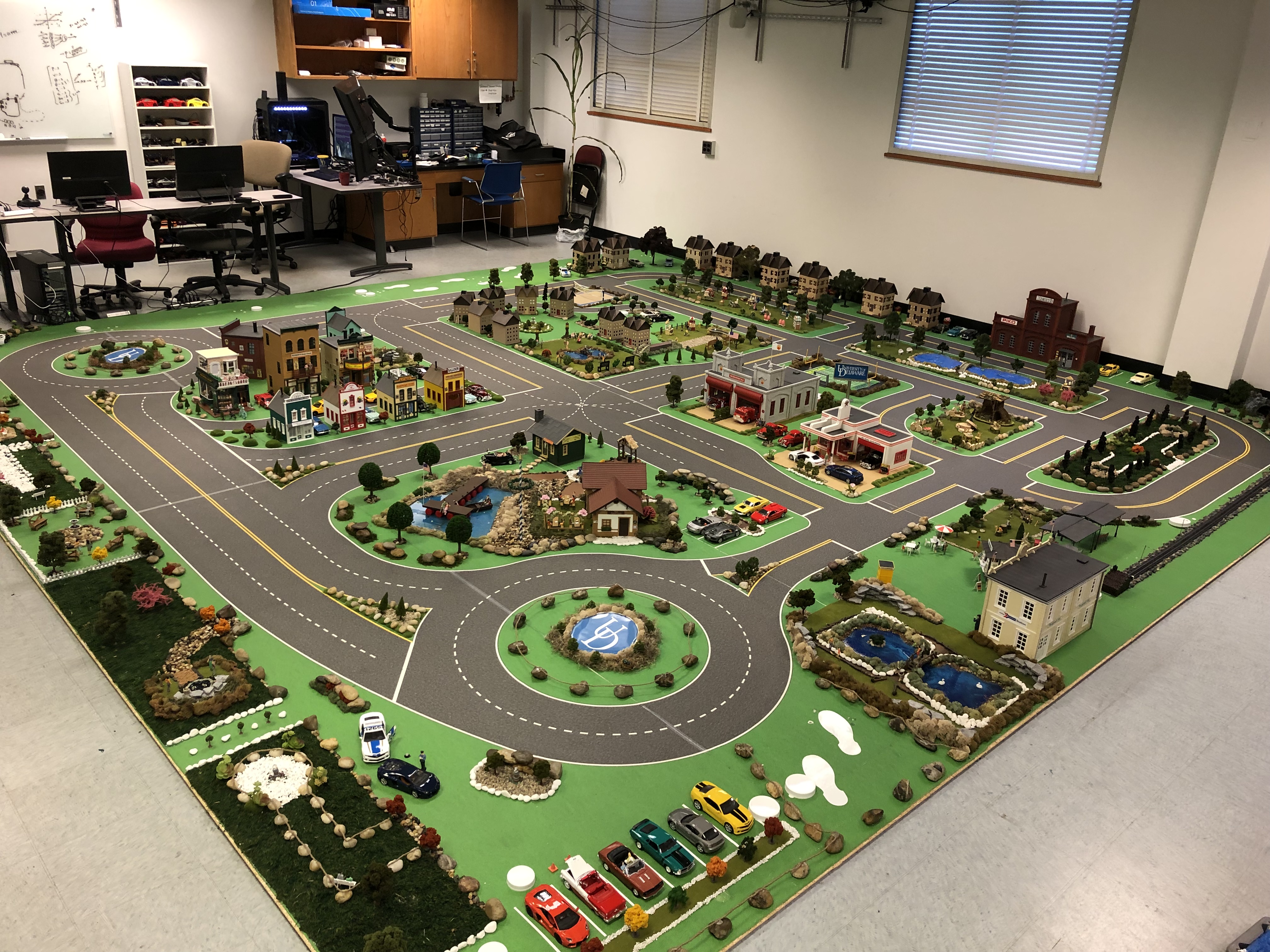}
    \caption{The University of Delaware's scaled smart city (UDSSC).}
    \label{fig:udssc}
\end{figure}}

\eat{\textcolor{gray}{
The scaled CAVs have been designed using easily assembled off-the-shelf components coupled with several 3D printed parts (Fig. \ref{fig:ScaledCAV}).
Each CAV consists of two primary boards, an Arduino Nano and a Raspberry Pi 3B with a $1.2$ GHz quad-core ARM processor and a $2.4$GHz WiFi adapter used to communicate with the central mainframe computer (Processor: Intel Core $i7-6950X$ CPU @ $3.00$ GHz x $20$, Memory: $125.8$ Gb). 
A power regulator manages the voltage input of the Raspberry Pi and Arduino, supplying a regulated $5$ VDC from two $3000$ mAh $3.7$ V Li-ion
batteries configured in series. With this hardware configuration, each CAV can run and collect experimental data at $20$ Hz for up to $2$ hours. 
Each CAV is rear-wheel drive with a pseudo-Ackerman steering system; the on-board Arduino provides the high-frequency control for the steering and motors. 
Each wheel is identical, with a rubber tire of radius $1.6~\text{cm}$.
}}

\eat{
\begin{figure}[ht]
    \centering
    \includegraphics[width=0.3\textwidth]{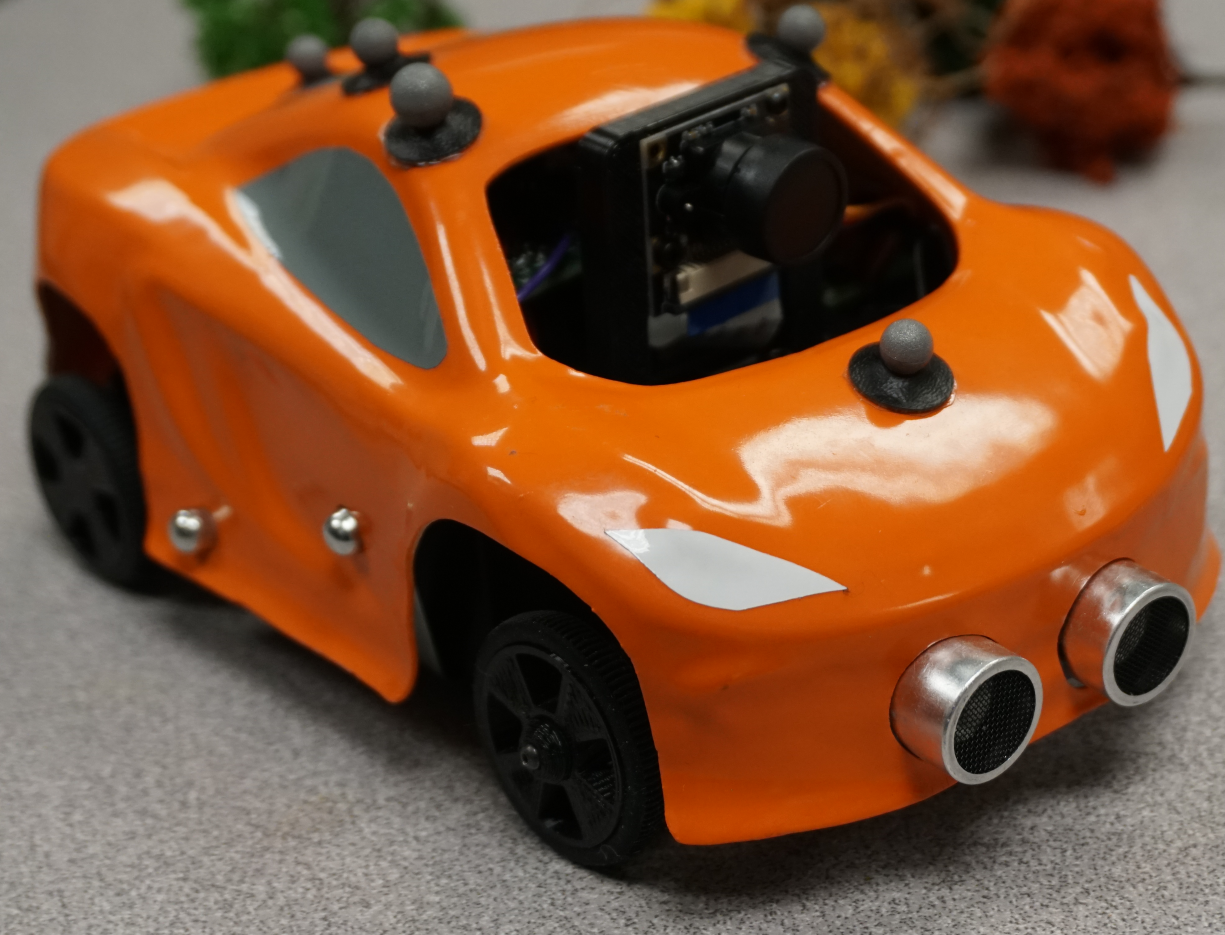}
    \caption{A picture of the connected and automated vehicles at the University of Delaware's scaled smart city (UDSSC).}
    \label{fig:ScaledCAV}
\end{figure}}

UDSSC utilizes a multi-level control architecture to precisely position each vehicle using position feedback from a VICON motion capture system. High-level routing and desired velocity calculation is handled by the mainframe computer, as well as locating the vehicle relative to each street on the map. This information is sent to each CAV which then calculates its desired steering and velocity actions based on a Stanley Controller \cite[eq. (9)]{Hoffman2017}, and
\eat{\textcolor{gray}{
The steering command $\delta(t)$ that is sent to each vehicle is calculated by
\begin{multline}
   \delta(t)=\Big(\Psi(t)-\Psi_{ss}(t)\Big)+\arctan{}\Big(\frac{k\,e(t)}{k_{soft}+v(t)}\Big)\\
   +k_{d,yaw}(r_{meas}-r_{traj})+k_{d,steer}\Big(\delta_{meas}(t - \Delta t)-\delta_{meas}(t)\Big),
    \label{eq:Stanley}
\end{multline}
as taken directly from \cite[eq.(9)]{Hoffman2017}. The parameters $k, k_{d,yaw}, k_{soft}, k_{d,steer}$ are all gains which are tuned for each vehicle; the values of $r_{meas}$ and $r_{traj}$ are the actual and desired yaw rates of the vehicles as calculated by VICON and the trajectory information, respectively, and $\Delta t$ is a time parameter. 
derivation and interpretation of the controller.
}}
the velocity control for each non-RL vehicle is specified by the IDM controller \eqref{eq:idm}.%; this is numerically integrated and sent to each vehicle as a velocity command.

\eat{
\textcolor{gray}{
Communication between the mainframe and each CAVs is achieved by a local WiFi network through a UDP/IP protocol connection at $50$ Hz. Each vehicle can measure its state of charge and can support additional sensors such as a Pi Camera or ultrasonic sensor.
}
}
%\subsection{Policy Transfer}
To implement the RL policy in UDSSC, the weights of the network generated by \textit{Flow} were exported into a data file. This file was accessed through a Python script on the mainframe, which uses a ROS service to map the current state of the experiment into a control action for each RL vehicle. During the experiment, the RL vehicles took commands from this script as opposed to the IDM controller.

To generate a disturbance on the roundabout system, a random delay for when each group was released was introduced. This delay was uniformly distributed between $0$ and $1$ seconds for the western group and between $0$ and $4$ seconds for the northern group during UDSSC experiments. The size of each vehicle group was randomly selected from a uniform distribution for each trial.

\subsection{Experimental Results}

The data for each vehicle was collected through the VICON motion capture system and is presented in Table \ref{tab:experimentalResults}. The position of each car was tracked for the duration of each experiment, and the velocity of each car was numerically derived with a first order finite difference method. 

\begin{table}[ht]
\centering 
   \caption{Experimental results for the baseline (no RL) case and each training method.}
    \begin{tabular}{lccccc} \label{tab:experimentalResults}
    Training & Mean    & Mean       & Trials & \% Time & Crashes\\
                  &Time (s) &Speed (m/s) &   & Saved &   \\\toprule
    Baseline    & 23.6 & 0.23 & 47 & - & 0 \\ \midrule
    \multicolumn{6}{c}{\textbf{Adversarial Multi-Agent}}\\
    Action-State & 22.1 & 0.24 & 29 & +6.3 & 0\\
    Action  & 22.1 & 0.23 & 23 & +6.4 & 0\\
    State   & 21.4 & 0.24 & 26 & +9.6& 10\\ \midrule
    \multicolumn{6}{c}{\textbf{Gaussian Single-Agent}}\\
    Action-State     & 25.8 & 0.21 & 26 & -9.2 & 0\\
    State    & 23.0 & 0.23 & 18 & +2.6 & 0\\
    Action   & 23.1 & 0.22 & 37 & +2.4 & 0\\
    Noiseless& 22.8 & 0.23 & 32 & +3.5 & 0
    \end{tabular}
\end{table}

\eat{
\begin{figure}[ht]
    \centering
    \includegraphics[width=0.45\textwidth]{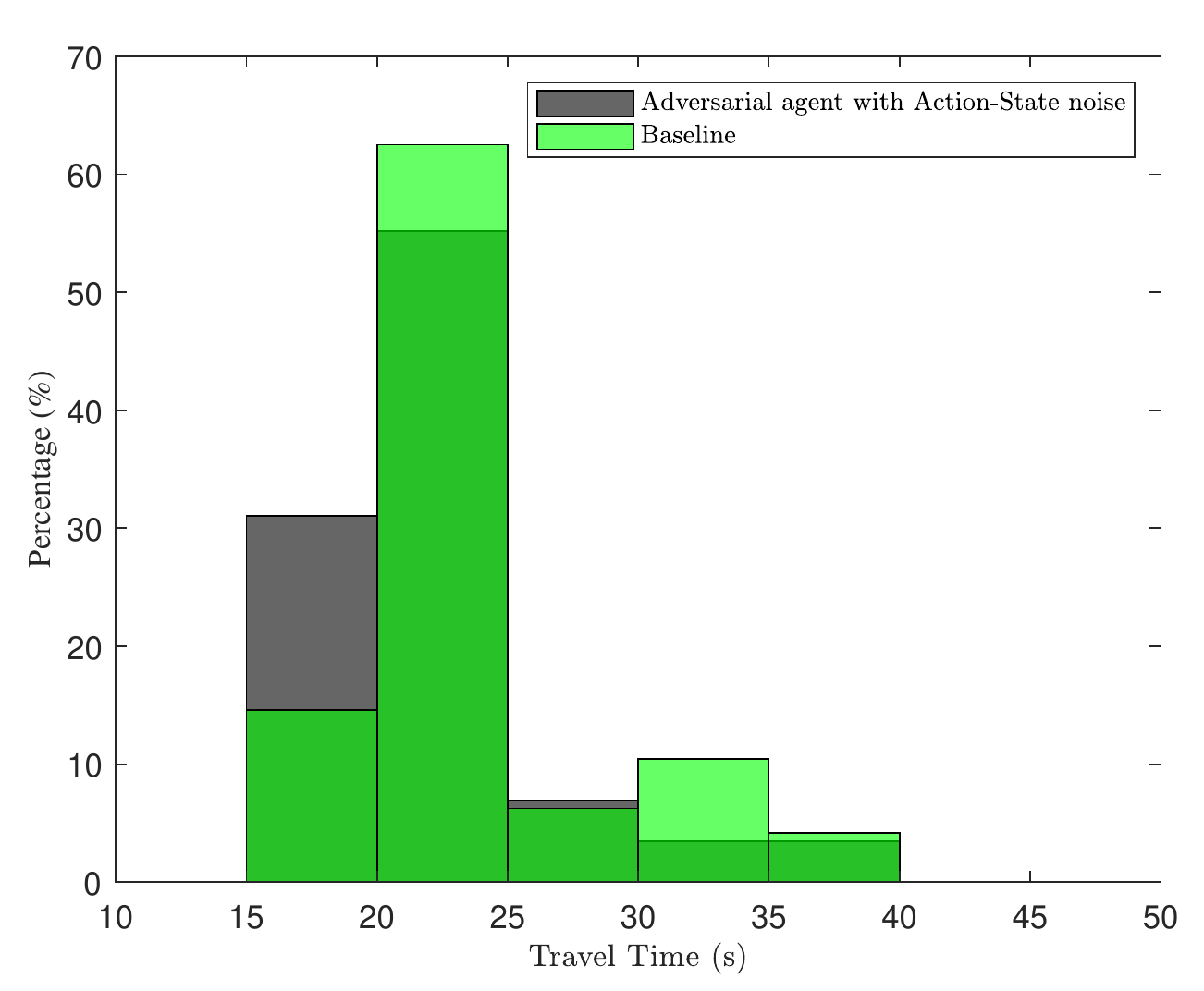}
    \caption{A relative frequency histogram for the travel time of each vehicle for the baseline and adversarial agent scenarios with noise injected in action and State.}
    \label{fig:TravelTimeHis}
\end{figure}

\begin{figure}[ht]
    \centering
    \includegraphics[width=0.42\textwidth]{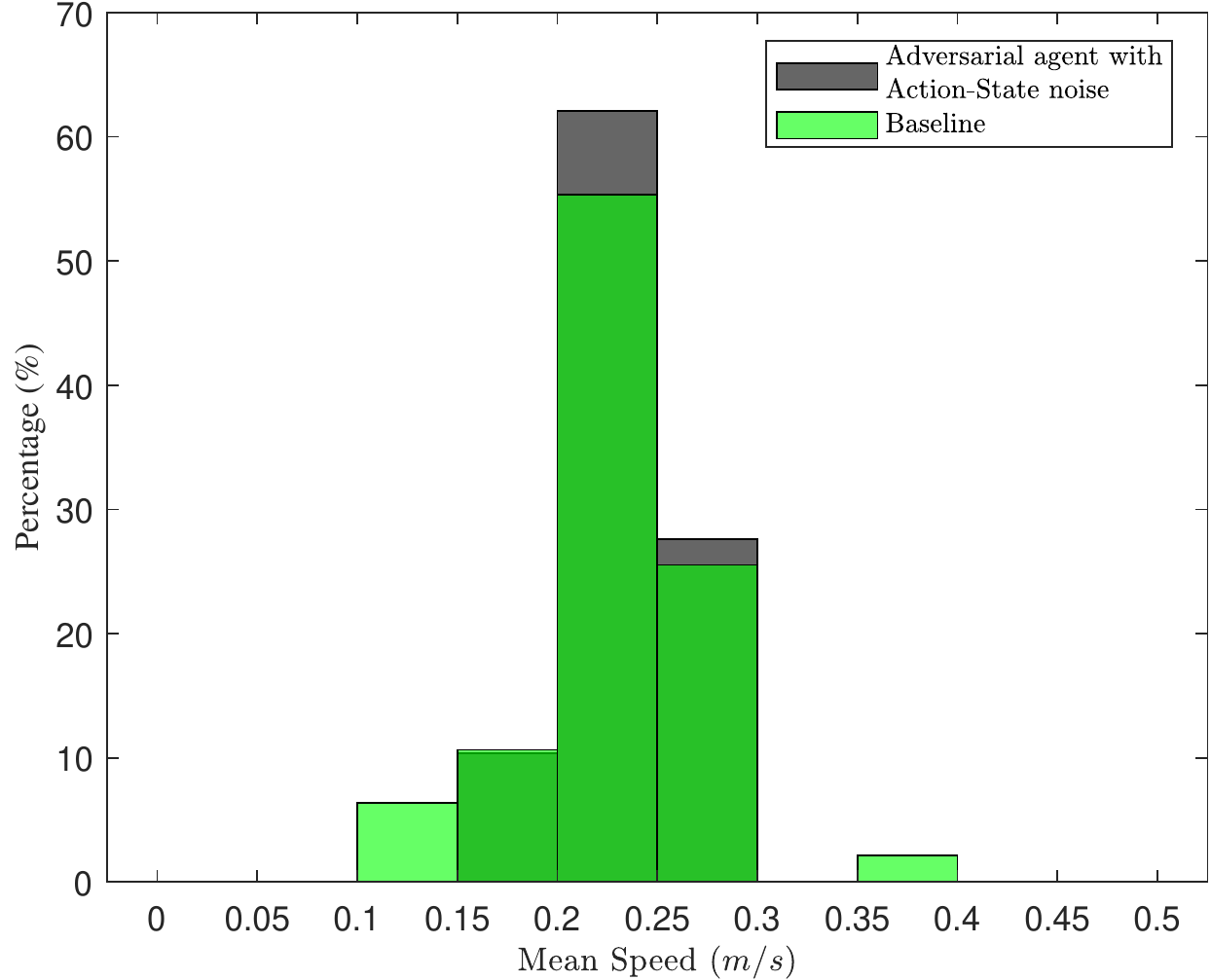}
    \caption{A relative frequency histogram of the mean speed of each vehicle for the baseline and adversarial agent scenarios with noise injected in action and state.}
    \label{fig:MeanSpeedHis}
\end{figure}
}

\begin{figure*}[ht]
    \centering
   \subfloat[][]{\includegraphics[width=0.35\linewidth]{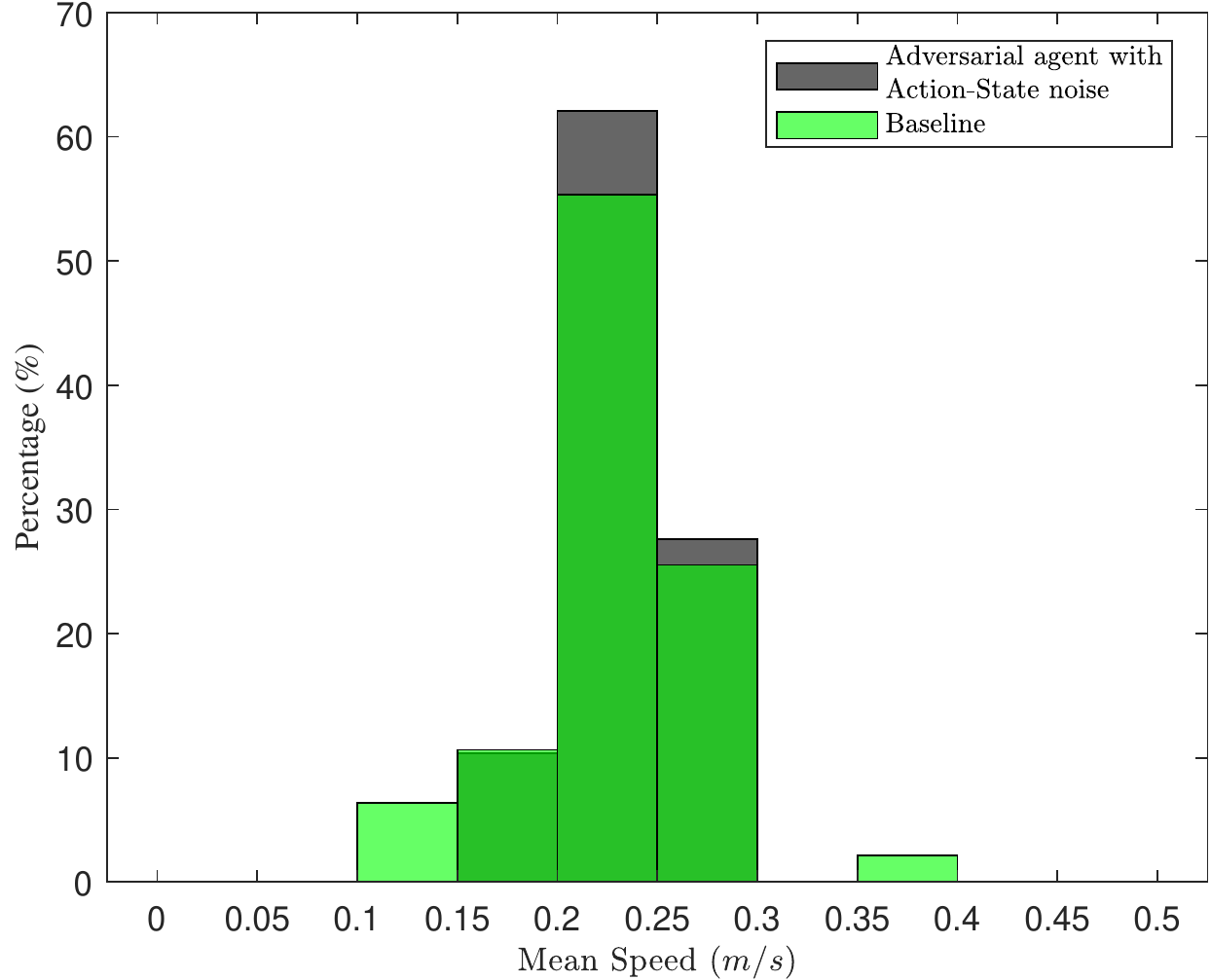}\label{a}}
     \subfloat[][]{\includegraphics[width=0.35\linewidth]{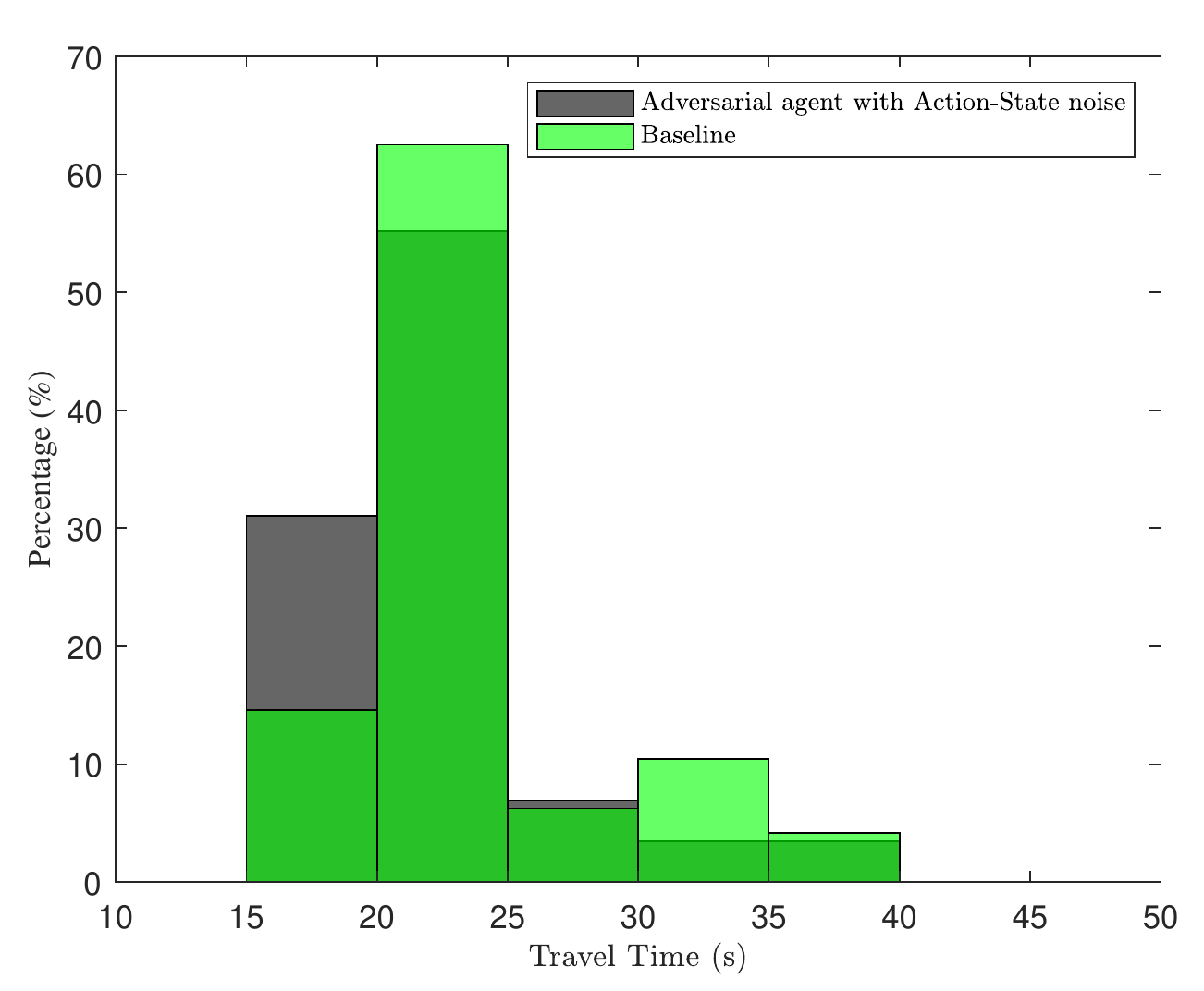}\label{b}}
    \caption{A relative frequency histogram for \protect\subref{a} the mean speed and \protect\subref{b} travel time of each vehicle for the baseline and adversarial multi-agent scenarios with noise injected in action and State.}
    \label{fig:Hist}
\end{figure*}

For all trials, the RL vehicle exhibited the learned ramp metering behavior, where the western leader reduced its speed to avoid yielding by the northern group. The metering behavior was extreme for the Gaussian single-agent noise case, especially when noise was added to the action and state together. This excessive metering significantly reduced the average speed and increased travel delay, as seen in Table \ref{tab:experimentalResults}. The adversarial multi-agent training significantly outperformed the Gaussian single-agent and tended to leave only a single vehicle yielding at the northern entrance. This strategy led to a travel time reduction for the northern group without a significant delay in western vehicles. The adversarial multi-agent case with noise injected only into the state accelerated especially fast and led to several catastrophic accidents between the two RL vehicles. Finally, for small numbers of vehicles, the adversarial multi-agent trained controllers appeared to exhibit an emergent zipper merging behavior.\footnote{ Videos of the experiment and supplemental information can be found at: \url{https://sites.google.com/view/ud-ids-lab/arlv}.}

Relative frequency histograms of average travel time and mean speed for the adversarial multi-agent case with noise injected into both action and state versus baseline scenarios are overlaid in Fig. \ref{fig:Hist}.
Over all trials, the adversarial case had a higher relative frequency of shorter travel time compared to the baseline scenario.
Average travel time for $30\%$ of adversarial %multi-agent with noise in action and state 
scenarios, lies in the range $[15 \text{s},20 \text{s}]$ comparing to $15\%$ for the baseline scenarios.

\eat{
\textcolor{gray}{
%Figure \ref{fig:Hist} depicts the relative frequency histogram of the mean speed for adversarial multi-agent with noise injected into both action and state scenarios overlaid on the baseline scenarios.
}}

From Table \ref{tab:experimentalResults}, we can see the average speed for baseline and the adversarial multi-agent with noise in action-state are nearly the same. However, in Fig. \ref{fig:Hist}, we can see that approximately $8\%$ of trials in the baseline scenarios have an average speed between $0.1$ m/s and $0.15$ m/s. Furthermore, there are some trials that the average speed of the baseline scenario is between $0.35$ m/s and $0.4$ m/s. On the other hand, the average speed for the adversarial multi-agent with noise in action and state varies less, and near $65\%$ of trials have the average speed between $0.2$ m/s and $0.25$ m/s compared to $55\%$ in the baseline scenarios.

\section{Conclusion} \label{sec:conclusions}

In this article, we developed a zero-shot transfer of an autonomous driving policy directly from simulation to the UDSSC testbed. Even under stochastic, real-world disturbances, the adversarial multi-agent policy improved system efficiency by reducing travel time and average speed for most vehicles. 
\eat{\textcolor{gray}{
Finally, we demonstrated that the addition of adversarial training in the state and noise of the policy considerably improves the stability and robustness of policies with respect to a real-world testbed during policy transfer. 
}}

As we continue to investigate approaches for policy transfer, some potential directions for future research include:
    multi-agent adversarial noise with multiple adversaries, 
    tuning to determine which elements of the state space are most suitable for perturbations,
    tuning injected noise to maximize policy robustness, 
    larger, more complex interactions, such as intersections, or merging at highway on-ramps, and
    longer tests involving corridors with multiple bottlenecks.
%\end{itemize}

\eat{\textcolor{blue}{
Another direction for future research should also include the generalization of the proposed framework to other traffic scenarios, i.e., intersections, speed reduction zones, including change lanes.
Furthermore, considering different penetrations of RL vehicles, which can significantly improve the efficiency of the entire network has to be studied.
}}

\section*{ACKNOWLEDGMENT}
The authors would like to thank Yiming Wan and Ishtiaque Mahbub for their contributions to UDSSC hardware and software.

\bibliographystyle{ieeetr} 
\bibliography{mybib.bib,UDSSC.bib,UDSSC2.bib}

\end{document}